\documentclass[%
 reprint,
nofootinbib,
 amsmath,amssymb,
 aps,
pra,
longbibliography]{revtex4-2}

\usepackage{graphicx}
\usepackage{dcolumn}
\usepackage{bm}

\usepackage{multirow}

\usepackage[margin=20truemm]{geometry}
\usepackage{tikz}
\usepackage{quantikz}
\usepackage{amsmath}
\usepackage{amssymb}
\usepackage{amsthm}
\usepackage{amsfonts}
\usepackage{graphicx}
\usepackage[colorlinks=true, allcolors=blue]{hyperref}
\usepackage{bbm}
\usepackage{complexity}
\usepackage{ifthen}
\usepackage[format=plain,justification=raggedright, singlelinecheck=false]{caption}

\newcommand{\cclasses}[1]{\begin{tikzpicture}
    \foreach \c[count=\n] in {#1}{\xdef\numclasses{\n}}
    \foreach \c[count=\n, evaluate=\n as \m using \numclasses-\n+1, evaluate=\n as \drk using 10+80*\n/\numclasses] in {#1}{
        \fill[gray!\drk] (0,\m/3) ellipse ({\xcontrol+.5*\m} and {\ycontrol+.5*\m});
        \ifthenelse{\n=1}{}{\node at (0,\m/3+\ycontrol+.5*\m){$\mayeq$};}
        \node[right,font={\sffamily}] at (0,{(\m/3+\ycontrol+.5*\m)-.45}){\c};}
    \draw ((0,\numclasses/3) ellipse ({\xcontrol+.5*\numclasses} and {\ycontrol+.5*\numclasses});
    \end{tikzpicture}}

\begin{document}

\title{
Hardness of classically sampling quantum chemistry circuits
}

\author{Ayoub Hafid,$^{1,2}$ Hokuto Iwakiri,$^2$ Kento Tsubouchi,$^3$ Nobuyuki Yoshioka,$^4$ and Masaya Kohda$^2$ }

\affiliation{\footnotesize $^1$\textit{Department of Mathematical Sciences, University of Tokyo, 3-8-1 Komaba, Tokyo 153-8914, Japan}}
\affiliation{\footnotesize $^2$\textit{QunaSys Inc., Aqua Hakusan Building 9F, 1-13-7 Hakusan, Bunkyo, Tokyo 113-0001, Japan}}
\affiliation{\footnotesize $^3$\textit{Department of Applied Physics, University of Tokyo,
7-3-1 Hongo, Bunkyo-ku, Tokyo 113-8656, Japan}}
\affiliation{\footnotesize  $^4$\textit{International Center for Elementary Particle Physics, University of Tokyo, Tokyo 113-0033, Japan}}

\def\prob#1{\mathrm{Prob}[#1]}
\newtheorem{lem}{Lemma}[section]
\newtheorem{thm}{Theorem}[section]
\newtheorem{prop}{Proposition}[section]
\newtheorem{define}{Definition}[section]
\newtheorem{exmp}{Example}[section]
\newtheorem{rmk}{Remark}[section]
\newtheorem{cor}{Corollary}[section]
\newtheorem{conj}{Conjecture}[section]

\begin{abstract}
Significant advances have been made in the study of quantum advantage both in theory and experiment, although these have mostly been limited to artificial setups.
In this work, we extend the scope to address quantum advantage in tasks relevant to chemistry and physics. Specifically, we consider the unitary cluster Jastrow (UCJ) ansatz---a variant of the unitary coupled cluster 
ansatz, which is widely used to solve the electronic structure problem on quantum computers---to show that sampling from the output distributions of quantum circuits implementing the UCJ ansatz is likely to be classically hard. More specifically, we show that there exist UCJ circuits for which classical simulation of sampling cannot be performed in polynomial time, under a reasonable complexity-theoretical assumption that the polynomial hierarchy does not collapse.
Our main contribution is to show that a class of UCJ circuits can be used to perform arbitrary instantaneous quantum polynomial-time (IQP) computations, which are already known to be classically hard to simulate under the same complexity assumption. As a side result, we also show that UCJ equipped with post-selection can generate the class post-BQP. Our demonstration, worst-case nonsimulatability of UCJ, would potentially imply quantum advantage in quantum algorithms for chemistry and physics using unitary coupled cluster type ansatzes, such as the variational quantum eigensolver and quantum-selected configuration interaction.
\end{abstract}

\maketitle

\section{Introduction}
\label{sec:intro}

Quantum algorithms are believed to be able to outperform classical counterparts in certain tasks. With recent progress in quantum devices, quantum advantage has been vigorously pursued, culminating in experimental demonstrations on quantum computers~\cite{google_rcs1,ustc_gbs1,xanadu,ustc_rcs2}.
A pressing challenge in this context is to expand the utility and advantage of using quantum computers beyond cryptographic purpose~\cite{rcsapplication,liu2025certified} to practical applications such as natural science.
In this regard, quantum chemistry and physics are promising targets for real-life applications of quantum computing.
While running long-term quantum algorithm with theoretical performance guarantee, namely the phase estimation~\cite{kitaev1995quantum}, requires fault tolerance for large-scale simulation, there are anticipations that near-term noisy quantum computers~\cite{nisq}, possibly with early-fault tolerance, could still provide quantum advantage in some tasks including dynamics simulation~\cite{utility} and a sampling-based configuration 
 interaction (CI) method~\cite{qsci}.

The sampling-based CI, or quantum-selected CI (QSCI), is a hybrid quantum-classical algorithm that employs quantum computers solely to sample from a many-body wave function to select Slater determinants for subspace diagonalization. 
Equipped with a novel post-processing technique, the method was demonstrated on a superconducting device for molecular systems up to 77 qubits~\cite{robledo2024chemistry}, which goes beyond the exact diagonalization, or full-CI method.
The method has attracted active research~\cite{nakagawa2024,kaliakin2024accurate,Barison_2025,liepuoniute2024quantum,shajan2024towards,sugisaki2024,mikkelsen2024quantum,reinholdt2025exposing,yu2025quantum,kaliakin2025implicit,yoshida2025auxiliary,danilov2025enhancing,barroca2025surfacereactionsimulationsbattery,shirai2025enhancing,ohgoe2025quantum}, both in pursuit of its practical applications and in efforts to better understand and improve its capabilities.
An underlying premise in the pursuit of quantum advantage, shared by this method as well as the well-known variational quantum eigensolver (VQE)~\cite{vqe}, is that sampling from certain quantum states of physical interest is classically intractable.
In this regard, a pressing open problem in the community is as follows: {\it Is there any quantum advantage in sampling task for quantum states considered in practical problems related to chemistry or physics?}

In this work, as a first step, we prove that there exist ansatz circuits, corresponding to  many-body wave functions of quantum chemistry and physics, that are hard to sample by classical computation, based on a reasonable complexity-theoretical assumption.
Specifically, we show the worst-case hardness of classically sampling the unitary cluster Jastrow (UCJ) ansatz~\cite{ucj}, which is a compact description of the renowned unitary coupled cluster ansatz~\cite{ucc}. This indicates that our result has implications for a wide class of quantum ansatzes for real applications.
Indeed, a hardware-efficient variant~\cite{motta2023bridging} of UCJ is adopted in the large scale demonstration~\cite{robledo2024chemistry} of QSCI.
In sharp contrast with the situation for local expectation values estimation tasks, where it remains an open question whether it is classically intractable for such classes of quantum circuits, our result establishes a firm foundation towards practical quantum advantage in the sampling-based CI algorithm.

The remainder of this paper is organized as follows.
In Sec.~\ref{sec:preliminaries}, we begin by recapitulating some basics of sampling problem and the existing proof~\cite{iqp2010} for classical hardness of sampling from instantaneous quantum polynomial-time (IQP) circuits.
This is because our proof relies on showing that any IQP circuit in a quadratic form (explained later), including those that are classically hard to sample, can be mapped to a quantum circuit representing the 1-UCJ ansatz, a single layer case of UCJ.
In Sec.~\ref{sec:results}, after introducing the UCJ ansatz, we show our proof for worst-case hardness of UCJ circuits.
We conclude with a summary and discussion on future directions in Sec.~\ref{sec:summary}. Some technical details are given in Appendix~\ref{appendix:diagonal}.

\section{Preliminaries
}\label{sec:preliminaries}

In this section, we recall some basics of sampling problem and how classical hardness of sampling IQP circuits can be shown based on some complexity-theoretical considerations. 

\subsection{
Sampling problem and classical simulatability
}\label{simulability_and_computational_complexity_classes}

Suppose a quantum circuit \( U \) acts on \( n \) qubits, initially prepared in a computational basis state, e.g., \(|0^n\rangle\). After applying \( U \), the system is in the state
\begin{align*}
|\psi\rangle = U|0^n\rangle.
\end{align*}
For such a state, measuring in the computational basis yields an outcome \( x \in \{0,1\}^n \), an $n$ bit string, with the probability
\begin{align*}
p(x) = |\langle x|\psi\rangle|^2.
\end{align*}

In this study, we consider classical simulation of such a sampling task for a certain class of quantum circuits.
Specifically, our study focuses on weak simulation under the multiplicative error criterion: a classical algorithm is supposed to sample from a distribution \( q(x) \) that approximates $p(x)$ as
\begin{align*}
\frac{1}{c}\, p(x) \le q(x) \le c\, p(x),
\end{align*}
where $c\geq 1$ indicates a multiplicative error.
We also focus on the worst-case hardness of classical simulation, that is, showing that there exists a circuit that cannot be efficiently simulated by classical computation.

Among the numerous studies on the classical hardness of sampling problem, particularly notable in the context of our work are those on the quantum approximate optimization algorithm~\cite{qaoa}, as well as on free-fermionic systems such as matchgate circuits~\cite{Valiant_2002,Jozsa_Miyake_2008a} and fermionic linear optics~\cite{flo} (see also Ref.~\cite{arrazola2022universal}).
Our proof relies on the results of Ref.~\cite{iqp2010}, where a class of circuits known as IQP is shown to be classically hard to sample in the worst case, based on a complexity-theoretical assumption about the polynomial hierarchy, which is an infinite tower of complexity classes.
As the authors of Ref.~\cite{iqp2010} point out, their results can be generalized to any class of circuits $\mathbf{A}$, with the main point rephrased as follows:
\begin{prop}\label{lem:postuniversalityandweaklyhardness}
If post-$\mathbf{A}$ $=$ post-BQP, then the weak classical simulatability of $\mathbf{A}$ within multiplicative error $1\leq c<\sqrt{2}$ would imply that the polynomial hierarchy collapses to the third level.
\end{prop}
Here, post-$\mathbf{A}$ denotes the class $\mathbf{A}$ augmented with post-selection on measurement outcomes. BQP can be viewed as a class of all quantum circuits consisting of a polynomial number of gates, and post-BQP refers to its extension with post-selection allowed.
As the polynomial hierarchy is widely believed not to collapse (see, e.g., Ref.~\cite{arora2009computational}), post-$\mathbf{A}$ $=$ post-BQP would imply that $\mathbf{A}$ is not weakly classically simulatable.

\subsection{
Nonsimulatability of IQP
}\label{IQP} 

\begin{figure}
        \begin{quantikz}
        \ket{0} & \gate[wires=1]{\mathcal{H}} &&  \gate[wires=3][2cm]{\mathcal{C}_{\rm diag}} && \gate{\mathcal{H}} & \meter{}\\
        \setwiretype{n}\phantom{Ca} \vdots \phantom{Ci} &   \phantom{Ca} \vdots \phantom{Ci}  && && \phantom{Ca} \vdots \phantom{Ci} & \vdots &&\\ 
        \ket{0} &  \gate[wires=1]{\mathcal{H}} &&                               && \gate{\mathcal{H}} & \meter{}\\
        \end{quantikz}
    \caption{Structure of IQP circuits. $\mathcal{H}$ denotes the Hadamard gate, and $\mathcal{C}_{\rm diag}$ is a diagonal circuit in the computational basis. In our study, we assume the form of $\mathcal{C}_{\rm diag}=e^{i\mathcal{D}}$, where $\mathcal{D}$ is a quadratic function of the Pauli operator $\mathcal{Z}$. Note that calligraphic letters are used to denote operators and gates for IQP circuits.
    }
    \label{fig:IQP_circuit}
\end{figure}
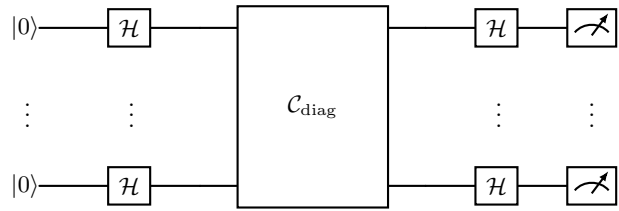

Here, we recapitulate the hardness proof~\cite{iqp2010} of classically sampling from the output distributions of IQP circuits.

IQP can be defined to be the class of circuits that have the following form: 
\begin{equation*}
    \mathcal{H}^{\otimes n} \mathcal{C}_{\rm diag} \mathcal{H}^{\otimes n},
\end{equation*}
where $n$ is the number of qubits, $\mathcal{H}$ is a Hadamard gate, and $\mathcal{C}_{\rm diag}$ denotes a circuit diagonal in the $Z$ basis.
We use calligraphic letters to denote operators and gates for IQP circuits throughout this paper.
The input state is $\ket{0^n}$, and the output is the measurement result in the computational basis.
The circuit diagram is shown in Fig.~\ref{fig:IQP_circuit}. 

IQP is a limited class of quantum circuits, but has the ability to perform universal quantum computation if post-selection is allowed. Indeed, one of the main results in Ref.~\cite{iqp2010} is the following:
\begin{prop}[a portion of Theorem 1 in Ref.~\cite{iqp2010}]\label{lem:post-iqp_post-bqp}
post-IQP =  post-BQP.
\end{prop}

To prove this, the following universal set of gates is considered in Ref.~\cite{iqp2010}: $H$, $Z$, $CZ$ and $e^{i\frac{\pi}{8}Z}$.
Among them, all gates except for the $H$ gate are diagonal in the $Z$ basis and, hence included in $\mathcal{C}_{\rm diag}$ of IQP.
Meanwhile, an arbitrary quantum circuit can be written in the form of $H^{\otimes n} C H^{\otimes n}$ using $H^2=I$, where $C$ consists of $H$ gates as well as the diagonal gates.
The authors show that each of the intermediate $H$ gates in $C$ can be removed by using a technique called Hadamard gadget when post-selection is allowed with an ancillary qubit, and an arbitrary circuit can be cast into the form of an IQP circuit augmented with post-selection. 
Therefore, post-IQP can perform universal quantum computation.

Now, Proposition~\ref{lem:postuniversalityandweaklyhardness} and \ref{lem:post-iqp_post-bqp} imply that the polynomial hierarchy would collapse to the third level if IQP could be weakly simulated.
Thus, given the collapse of the polynomial hierarchy is unlikely, it is implausible that IQP can be efficiently sampled by classical computation with multiplicative error $1 \leq c < \sqrt{2}$.

In our study, we assume the diagonal circuit to take the restricted form of $\mathcal{C}_{\rm diag}=e^{i\mathcal{D}}$ with
\begin{align}\label{eq:IQP-D-term}
\mathcal{D} = \sum_{\alpha <\beta}w_{\alpha\beta}\mathcal{Z}_\alpha \mathcal{Z}_\beta +\sum_{\alpha}v_\alpha \mathcal{Z}_\alpha,
\end{align}
where $\alpha,\beta$ are qubit indices, and $w_{\alpha\beta}, v_\alpha$ are real parameters.
Nonetheless, the complexity argument is not affected as $e^{i\mathcal{D}}$ is capable of implementing $e^{i\theta \mathcal{Z}}$ and $e^{i\theta (\mathcal{I}-\mathcal{Z})\otimes (\mathcal{I}-\mathcal{Z})}$ gates, which in turn can implement any diagonal gate up to a global phase in the universal gate set adopted in the proof.

We were led to our proof by observing that this form of IQP circuit, $\mathcal{H}^{\otimes n}e^{i\mathcal{D}}\mathcal{H}^{\otimes n}$, is reminiscent of the 1-UCJ ansatz $e^{-\hat{K}}e^{\hat{J}}e^{\hat{K}}$ (explained later).
We flesh out this intuition to show that any IQP circuit with $\mathcal{D}$ in this quadratic form can be emulated by some 1-UCJ circuit.
Hence, the classical hardness of 1-UCJ is inherited from IQP.

\section{Main results}
\label{sec:results}

We now present our main results, namely the worst-case hardness of sampling from quantum-chemistry circuits by classical computation.
Specifically, we consider the UCJ ansatz~\cite{ucj}, designed as a compact description of electronic structure on a quantum computer.
Our argument heavily draws on the hardness proof of IQP~\cite{iqp2010}.
What we prove here is that an arbitrary IQP circuit of $n$ qubits can be efficiently emulated by a UCJ circuit of single layer (1-UCJ) with $4n$ qubits, or $4n$ spin orbitals.
Assuming there are IQP circuits that are classically hard to sample from, this in turn implies the existence of 1-UCJ circuits with the same hardness. We focus on the noiseless simulation.

\begin{figure}[b]
\includegraphics[scale=0.16]{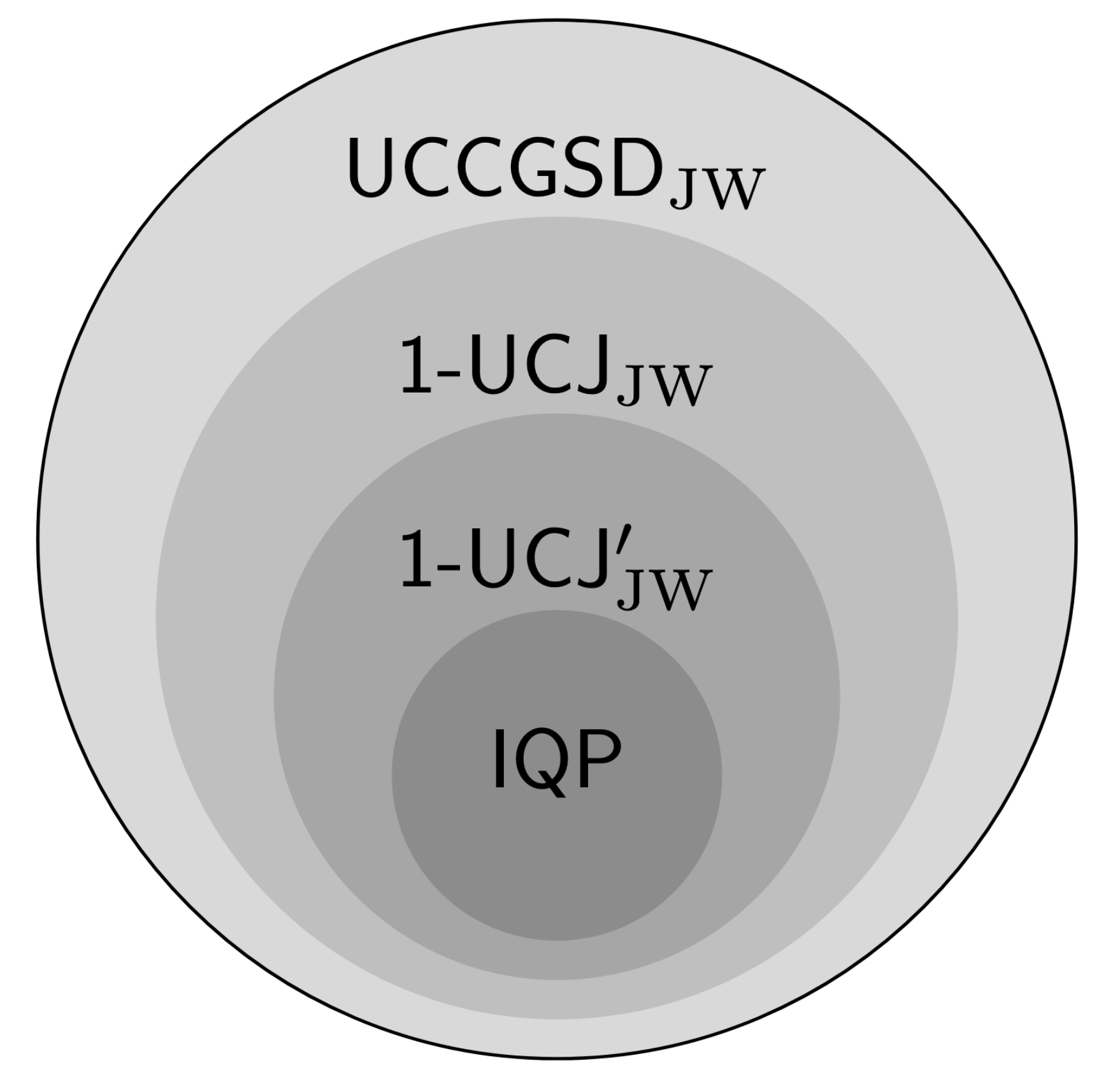}
\caption{
Complexity hierarchy of the classes of quantum circuits considered in this work. UCCGSD$_{\rm JW}$ stands for the class of quantum circuits obtained from the unitary coupled cluster ansatz with generalized singles and doubles, via the JW transformation. 1-UCJ$_{\rm JW}$ is similarly defined from the UCJ ansatz with a single layer. 1-UCJ$_{\rm JW}^\prime$ is a restricted set of 1-UCJ$_{\rm JW}$ as defined in Eq.~\eqref{eq:1-ucj-prime-def}.
IQP circuits are assumed to take the form of $\mathcal{H}^{\otimes n}e^{i\mathcal{D}}\mathcal{H}^{\otimes n}$ with $\mathcal{D}$ defined as the quadratic form of Eq.~\eqref{eq:IQP-D-term}.
}
\label{fig:hierarchy}
\end{figure}

Our actual proof introduces two more families of circuits, 1-UCJ$_{\rm JW}$ and 1-UCJ$_{\rm JW}^\prime$.
Here, 1-UCJ$_{\rm JW}$ denotes a set of circuits that are obtained from 1-UCJ ansatz states mapped to quantum circuits via the Jordan--Wigner (JW) transformation.
1-UCJ$_{\rm JW}^\prime$ refers to a restricted set of 1-UCJ$_{\rm JW}$, and is introduced in intermediate steps of the proof.
We then show that the class 1-UCJ$_{\rm JW}^\prime$ contains IQP, implying:
\begin{thm}\label{th:our-theorem}
1$\mathchar`-$UCJ$_{JW}$ $\supseteq$ IQP.
\end{thm}
Then, the classical hardness of 1-UCJ is implied by the IQP result.
Figure~\ref{fig:hierarchy} illustrates the inclusion relations of the circuit classes considered in this paper.

After briefly reviewing the UCJ ansatz, the following subsections describe our proof. We defer some technical details to Appendix~\ref{appendix:diagonal}.

%

\subsection{
1-UCJ ansatz
}
\label{ucj-representation-subsection}

In this study, we consider the UCJ ansatz~\cite{ucj} to investigate computational complexity. It is a unitary variant of the cluster Jastrow ansatz~\cite{neuscamman2013}, and can be viewed as a compact description of the unitary coupled cluster ansatz with generalized singles and doubles~\cite{lee2018generalized}, while still maintaining accuracy.

In general, a UCJ ansatz consists of $k$ repetitions of a triple product of exponential operators in the form of $e^{-\hat{K}}e^{\hat{J}}e^{\hat{K}}$, which is dubbed the $k$-UCJ ansatz.
In this study, we focus on the single-layer case, namely, the 1-UCJ ansatz defined as 
\begin{equation}\label{eq:ucj-def}
U_{\textrm{1-UCJ}}
= e^{-\hat{K}}e^{\hat{J}}e^{\hat{K}},
\end{equation}
where\footnote{
We note some notational differences from the original work~\cite{ucj}.
There is a sign difference preceding $\hat{K}$ in Eq.~\eqref{eq:ucj-def}.
Besides, we adopt a common orbital basis for the definitions of both $\hat{J}$ and $\hat{K}$.
Nonetheless, our definition is equivalent to theirs, given appropriate redefinitions of the parameters.
}
\begin{align}\label{eq:J-K}
\hat{J}&=\sum_{p, q, \sigma, \tau} J_{pq}^{(\sigma\tau)} a_{p\sigma}^{\dag}a_{p\sigma}a_{q\tau}^{\dag}a_{q\tau}, \notag \\
\hat{K} &= \sum_{p,q,\sigma} K_{pq} a_{p\sigma}^{\dag}a_{q\sigma}.
\end{align}
Here, $p,q$ denote molecular orbitals while $\sigma,\tau = \uparrow,\downarrow$ are spin indices.
$J_{pq}^{(\sigma\tau)}$ and $K_{pq}$ are parameters of the ansatz.
$K_{pq}$ constitute an anti-Hermitian matrix with respect to indices $p$ and $q$, while $J_{pq}^{(\sigma\tau)}$ are purely imaginary parameters satisfying $J_{pq}^{(\sigma\tau)} = J_{qp}^{(\sigma\tau)}$ and $J_{pq}^{(\sigma\tau)} = J_{pq}^{(\tau\sigma)}$.
These conditions are sufficient to ensure unitarity of the ansatz.
$J_{pq}^{(\uparrow\uparrow)}=J_{pq}^{(\downarrow\downarrow)}$ is further assumed in Ref.~\cite{ucj}, but our proof holds regardless of this condition.
In our proof, we impose $J_{pq}^{(\uparrow\downarrow)} = J_{pq}^{(\downarrow\uparrow)} = 0$ and $K_{pq}\in \mathbb{R}$.

The wave function of electrons for the UCJ ansatz is given as $U_{\textrm{1-UCJ}}\ket{\psi_{\rm ref}}$, where $\ket{\psi_{\rm ref}}$ is a reference state.
In our study, the Hartree--Fock (HF) state is taken as the reference state.
Note that any 1-UCJ ansatz is a special case of the unitary coupled cluster ansatz with generalized singles and doubles.

\subsection{
1-UCJ$_{\rm JW}^\prime$: a restricted class of 1-UCJ circuits with JW transformation
}

The 1-UCJ ansatz $e^{-\hat{K}}e^{\hat{J}}e^{\hat{K}}$, and $k$-UCJ in general, can be implemented as a quantum circuit without relying on Trotter decomposition.
This is because each term in $\hat{J}$, $a_{p\sigma}^{\dag}a_{p\sigma}a_{q\tau}^{\dag}a_{q\tau}$, is a product of number operators, which commute with each other, and the orbital rotations $e^{\hat{K}}$ can be implemented using Givens rotations~\cite{wecker2015solving,decompconnectivity}.

In this study, we adopt the JW transformation to map fermionic operators into qubit representations.
We define the class 1-UCJ$_{\rm JW}$ as a collection of quantum circuits obtained from the 1-UCJ ansatz via the JW transformation.
We impose additional constraints to define a restricted class of 1-UCJ$_{\rm JW}$, namely 1-UCJ$_{\rm JW}^\prime$.
We show that each IQP circuit can be mapped to a specific 1-UCJ$_{\rm JW}^\prime$ circuit.

To simplify the proof, we focus on the spin-up sector by decoupling the spin-down sector from the system with the assumption $J_{pq}^{(\uparrow\downarrow)} = J_{pq}^{(\downarrow\uparrow)} = 0$.
Under this assumption, the UCJ ansatz state can now be written in the form of
\begin{align*}
&U_{\textrm{1-UCJ}}\ket{\psi_{\rm ref}} \\
&= \left( e^{-\hat{K}_\uparrow}e^{\hat{J}_\uparrow}e^{\hat{K}_\uparrow}\ket{\psi_{\rm ref}^{(\uparrow)}} \right) \otimes \left( e^{-\hat{K}_\downarrow}e^{\hat{J}_\downarrow}e^{\hat{K}_\downarrow}\ket{\psi_{\rm ref}^{(\downarrow)}} \right).
\end{align*}
Here, the reference state, which we assume to be the HF state, is separated into spin-up and spin-down sectors as $\ket{\psi_{\rm ref}} = \ket{\psi_{\rm ref}^{(\uparrow)}}\otimes \ket{\psi_{\rm ref}^{(\downarrow)}}$, and $\hat{J}$ and $\hat{K}$ are divided into spin-specific terms as $\hat{J}_\sigma =\sum_{ p, q} J_{pq}^{(\sigma\sigma)} a_{p\sigma}^{\dag}a_{p\sigma}a_{q\sigma}^{\dag}a_{q\sigma}$ and $\hat{K}_\sigma = \sum_{p,q} K_{pq} a_{p\sigma}^{\dag}a_{q\sigma}$ for $\sigma=\uparrow, \downarrow$, respectively.
After tracing out the spin-down sector, the subsystem of spin-up electrons is described by a pure state in the form of $e^{-\hat{K}_\uparrow}e^{\hat{J}_\uparrow}e^{\hat{K}_\uparrow}\ket{\psi_{\rm ref}^{(\uparrow)}}$. In the following, we ignore the spin-down sector and omit the spin indices $\sigma, \tau$.

We further assume $K_{pq}$ is a real anti-symmetric matrix and rewrite the orbital rotation as
\begin{align}\label{eq:orbital-rot}
    e^{\hat{K}}
    &= \exp\left[\sum_{p < q}     K_{pq}\left( a_{p}^{\dagger}\,a_{q} 
        - a_{q}^{\dagger}\,a_{p} \right)
    \right].
\end{align}
Using Givens rotations~\cite{wecker2015solving,decompconnectivity}, this can be expressed as a sequence of rotations $R_{pq}$, each acting on a pair of spin orbitals, given by
\begin{align}\label{eq:R_pq}
R_{pq}(T_{pq}) = \exp\left[ T_{pq} \left( a_{p}^{\dagger}a_{q} -a_{q}^{\dagger}a_{p} \right)\right].
\end{align}
Specifically, $e^{\hat{K}}$ can be decomposed as 
\begin{align*}
    &e^{\hat{K}}
    = \prod_{p < q} R_{pq}(T_{pq}),
\end{align*}
for some real parameters $T_{pq}$.
Conversely, any such product of $R_{pq}(T_{pq})$, with arbitrary real parameters $T_{pq}$, can be written as $e^{\hat{K}}=\exp[\sum_{p<q}K_{pq}(a_p^\dagger a_q -a_q^\dagger a_p)]$ for some real parameters $K_{pq}$. This property plays a key role in relating arbitrary IQP circuits to 1-UCJ counterparts.

The operator $e^{\hat{J}}$, which consists of number operators, can be straightforwardly decomposed as
\begin{align*}
e^{\hat{J}}
= \exp\left( \sum_{p\leq q}J_{pq}a_p^\dagger a_p a_q^\dagger a_q \right)
= \prod_{p \leq q}
    D_{pq}(-i J_{pq}),
\end{align*}
where we define
\begin{align}\label{eq:D_pq}
D_{pq}(\theta) = \exp\left( i\theta a_{p}^{\dagger} a_{p} a_{q}^{\dagger} a_{q} \right),
\end{align}
for a real number $\theta$.
Note $J_{pq}$ is pure imaginary.
We restrict the range of $p,q$ to $p\leq q$ by leveraging the symmetry $J_{pq}=J_{qp}$ with a suitable rescaling of $J_{pq}$.

Concentrating on the spin-up sector, then, the 1-UCJ ansatz defined in Eq.~\eqref{eq:ucj-def} can be expressed as
\begin{align*}
&U_{\textrm{1-UCJ}} \\
&=\left(\prod_{p<q}R_{pq}(T_{pq})\right)^{\dagger}
\left(\prod_{p\leq q}D_{pq}(\theta_{pq})\right)
\left(\prod_{p < q}R_{pq}(T_{pq})\right),
\end{align*}
where $\theta_{pq} = -i J_{pq} \in \mathbb{R}$.

We now apply the JW transformation to the fermionic system of $2n$ spin-up orbitals. We adopt the following ordering for qubits (orbitals):
\begin{align*}
\ket{b_0 b_1 \cdots b_{2n-1}},
\end{align*}
which corresponds to a computational basis state (Slater determinant) with $b_p \in \{0,1\}$.
For simplicity, we assume half of the $2n$ orbitals are occupied in the reference state, which we take as
\begin{align}\label{eq:ref-state}
\ket{\psi_{\rm ref}}
= \ket{0_0 \cdots 0_{n-1} 1_n \cdots 1_{2n-1}}.
\end{align}
Extending to cases where the half-filling condition is not satisfied is straightforward.

The fermionic operators appearing in $R_{pq}$ and $D_{pq}$ are written as linear combinations of Pauli strings.
For instance, $a_p^\dagger a_q$ is expressed for $p < q$ as
\begin{align*}
&a_p^\dagger a_q \\
&= \frac{X_p - iY_p}{2}\otimes Z_{p+1}\otimes \cdots \otimes Z_{q-1}\otimes \frac{X_q + iY_q}{2}.
\end{align*}
The number operator for $p$-th orbital is given as $a_p^\dagger a_p = (I_p-Z_p)/2$. Hence, $D_{pq}$ is expressed for $p\neq q$ as
\begin{align}\label{eq:D_pq_JW}
D_{pq}(\theta) = \exp\left[ i\theta \left(\frac{I_p -Z_p}{2}\right)\otimes\left(\frac{I_q -Z_q}{2}\right) \right],
\end{align}
while $D_{pp}(\theta) = \exp[i\theta (I_p - Z_p)/2]$.
Their matrix representations are given as
\begin{align}
D_{pq}(\theta) =
\begin{pmatrix}
    1&0&0&0\\
    0& 1 & 0 &0\\
    0& 0 & 1 &0\\
    0&0&0& e^{i\theta}
\end{pmatrix}, \notag
\end{align}
for $(\ket{0 0}, \ket{0 1}, \ket{1 0}, \ket{1 1})$, and
\begin{align*}
D_{pp}(\theta) =
\begin{pmatrix}
    1 & 0 \\
    0 & e^{i\theta}
\end{pmatrix},
\end{align*}
for $(\ket{0},\ket{1})$.
Thus, $D_{pp}$ gives the phase shift gate, while $D_{pq}$ ($p\neq q$) the controlled phase shift gate.

Furthermore, $R_{pq}(\theta)=\exp[\theta (a_p^\dagger a_q -a_q^\dagger a_p)]$ acts on qubits $p$ through $q$.
Its JW representation corresponds to the Givens rotation that mixes $\ket{0_p}\ket{1_q}$ and $\ket{1_p}\ket{0_q}$, up to a $Z$ string.
For a descriptive purpose, we first consider the matrix representation of $R_{pq}$ on a Hilbert subspace where $(p+1)$-th to $(q-1)$-th qubits, for $q>p+1$,\footnote{Note that we only need to consider the cases where $q>p$. When $q=p+1$, the JW representation of $R_{pq}$ reduces to the Givens rotation without the parity factor $(-1)^{w(b)}$.}  are fixed as the computational basis state $\ket{b}=\ket{b_{p+1}\cdots b_{q-1}}$ with $b_r \in \{0,1\}$.
Then, the matrix representation of $R_{pq}$ is given as
\begin{align}\label{eq:R_pq_JW}
R_{pq}(\theta) =
\begin{pmatrix}
    1&0&0&0\\
    0& \cos\theta & (-1)^{w(b)+1}\sin\theta &0\\
    0& (-1)^{w(b)}\sin\theta & \cos\theta &0\\
    0&0&0&1
\end{pmatrix},
\end{align}
for $(\ket{0_p b  0_q}, \ket{0_p b 1_q}, \ket{1_p b 0_q}, \ket{1_p b 1_q})$.
Here, $w(b)$ denotes the Hamming weight of the bit string $b$, that is, the number of ones in $b$, and the factor $(-1)^{w(b)}$ gives the parity of $b$.
It is straightforward to see that the matrix expression is similarly given for less restrictive subspace where the parity of Hamming weights for intervening qubits are fixed.
This condition is satisfied throughout the remainder of the proof.

We observe that $R_{pq}$ acts nontrivially only on the two-dimensional subspace. If we identify $(\ket{0_p}\ket{1_q}, \ket{1_p}\ket{0_q})$ as an effective single qubit, then $R_{pq}((-1)^{w(b)}\theta/2)$ corresponds to the single-qubit rotation about the $y$-axis:
\begin{align*}
\mathcal{R}_y(\theta) = e^{-i\theta \mathcal{Y}/2}
=\begin{pmatrix}
    \cos\frac{\theta}{2} & -\sin\frac{\theta}{2} \\
    \sin\frac{\theta}{2} & \cos\frac{\theta}{2}
\end{pmatrix}.
\end{align*}
Here and in what follows, calligraphic letters are used to denote operators and gates acting on effective qubits, anticipating their identification with those of IQP circuits.
Note that the angle in $R_{pq}$ is redefined to cancel the parity factor $(-1)^{w(b)}$ in Eq.~\eqref{eq:R_pq_JW}.

\begin{figure*}[tb]
\includegraphics[width=0.8\textwidth]{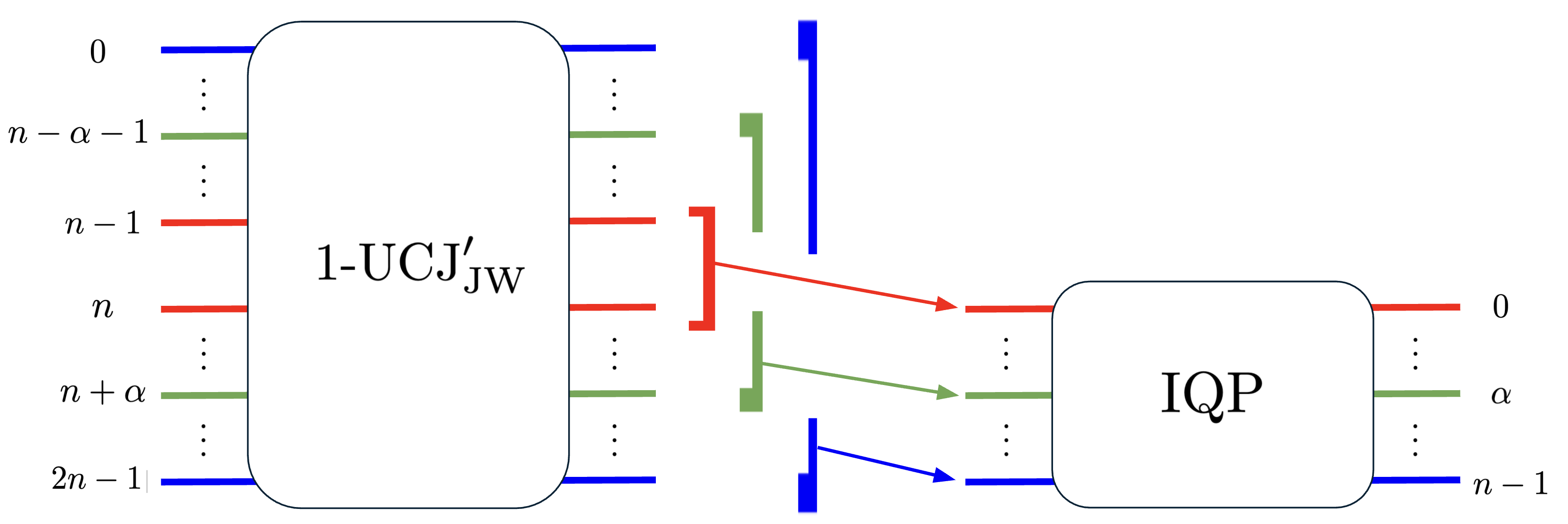}
    \caption{Identification of 1-UCJ qubits with IQP qubits. Specifically, 1-UCJ$_{\rm JW}^\prime$, a restricted version of 1-UCJ with JW transformation, is considered. For each specific pair of qubits in 1-UCJ, the single particle subspace spanned by $(\ket{01}, \ket{10})$ is identified as a single qubit in IQP. This correspondence is expressed as $\ket{\bar{0}}=\ket{01}$, $\ket{\bar{1}}=\ket{10}$.
    In the chemist's language, we pair up HOMO$-\alpha$ with LUMO$+\alpha$ for $\alpha=0,1,\cdots, n-1$ in the HF state of Eq.~\eqref{eq:ref-state}. Here, HOMO and LUMO stand for the highest occupied molecular orbital and the lowest unoccupied molecular orbital, respectively.
    }
    \label{fig:1-UCJ_IQP}
\end{figure*}
\begin{figure*}[tb]
    \scalebox{0.8}{
        \begin{quantikz}
        \ket{0} && \gate[wires=8]{ V_{n-1}} && &&  && && \gate[wires=8][1cm]{\prod_{p\le q} D_{pq}(\theta_{pq})} && && && && \gate[wires=8]{V^{\dagger}_{n-1}} &&\\
        \ket{0} &&                         && \gate[wires=6]{V_{n-2}}  &&  && &&                               && 
         && && \gate[wires=6]{V^{\dagger}_{n-2}} && &&\\
        \phantom{Ca} \vdots \phantom{Ci}\\ 
        \ket{0} && &&                   && \ \ldots \ &&  \gate[wires=2]{ V_{0}} &&          && \gate[wires=2]{V^{\dagger}_{0}} && \ \ldots \     && && &&\\
        \ket{1} &&   &&                       && \ \ldots \ && &&                               && 
         &&   \ \ldots \ &&  && &&\\
        \phantom{Ca} \vdots \phantom{Ci}\\ 
        \ket{1} &&  &&                     &&  &&  &&                             && 
         &&  &&  && &&\\ 
        \ket{1} &&                        && && &&     &&                          && && && &&  &&\\
        \end{quantikz}
    }
    \caption{A 1-UCJ$_{\rm JW}^\prime$ circuit with $2n$ qubits, defined in Eq.~\eqref{eq:1-ucj-prime-def}.
    This is a 1-UCJ ansatz with $2n$ orbitals mapped to qubits by the JW transformation, with parameters of the orbital rotation $e^{\hat{K}}$ restricted. The gate
    $V_\alpha$ and $D_{pq}(\theta)$ are defined in Eqs.~\eqref{eq:gate-V},~\eqref{eq:D_pq_JW}, respectively.
    }
    \label{fig:1-UCJ-prime}
\end{figure*}
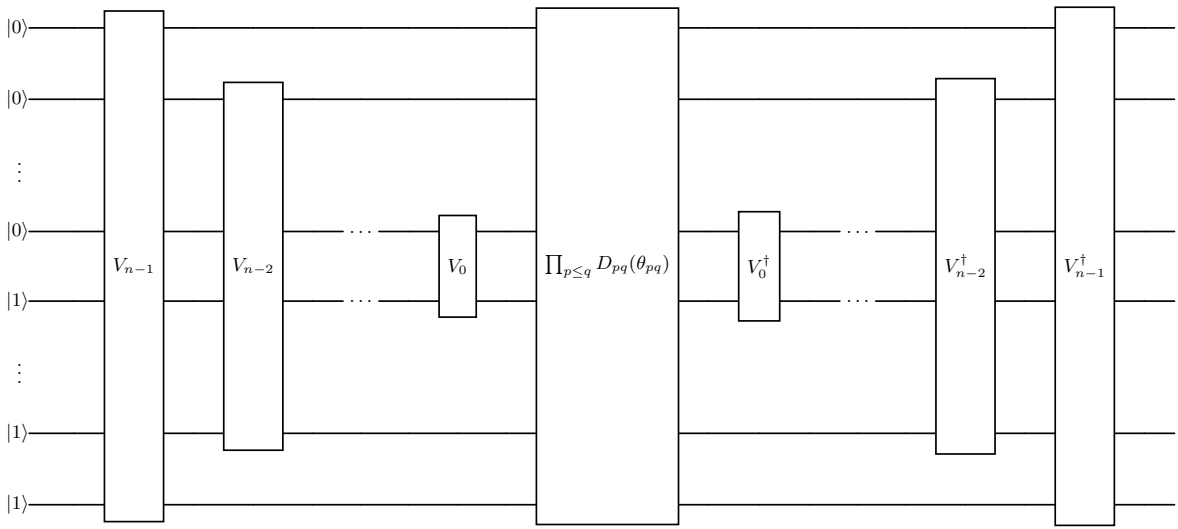

We utilize this identification to associate an $n$-qubit IQP circuit with a $2n$-orbital 1-UCJ ansatz. As illustrated in Fig.~\ref{fig:1-UCJ_IQP}, we pair one occupied orbital with one unoccupied orbital in the HF state of Eq.~\eqref{eq:ref-state}, and refer to the effective qubit as IQP qubit.
Recall that an IQP circuit takes the form as\begin{align*}
\mathcal{H}^{\otimes n} e^{i \mathcal{D}} \mathcal{H}^{\otimes n}.
\end{align*}
Noting the relation $\mathcal{R}_y(-\pi/2) = \mathcal{ZH}$, we construct the Hadamard gate $\mathcal{H}$ of IQP from an orbital rotation $R_{pq}((-1)^{w(b)+1}\pi/4)$.
We show that the diagonal part $e^{i\mathcal{D}}$ can be constructed by an appropriate sequence of $D_{pq}$, which is also diagonal.

We now introduce 1-UCJ$_{\rm JW}^\prime$, a restricted class of JW-transformed 1-UCJ circuits on $2n$ qubits.
With the reference state of Eq.~\eqref{eq:ref-state}, it is defined by fixing the parameters of $R_{pq}$:
\begin{align}\label{eq:1-ucj-prime-def}
&U_{\textrm{1-UCJ}_{\rm JW}^\prime} \notag\\
&=V^{\dagger}_{n-1} \cdots V^{\dagger}_{0} \left(\prod_{p\leq q} D_{pq}(\theta_{pq}) \right) V_{0} \cdots V_{n-1},
\end{align}
where $V_\alpha$ corresponds to $R_{pq}(T_{pq})$ with parameters fixed as $T_{pq} = (-1)^{q-n+1}\frac{\pi}{4} \delta_{p+q,2n-1}$, yielding
\begin{align}\label{eq:gate-V}
V_\alpha = R_{n-\alpha -1,n+\alpha}\left((-1)^{\alpha +1}\frac{\pi}{4}\right),
\end{align}
for $\alpha =0,1,\cdots, n-1$.
No additional restriction is introduced for $\prod_{p\leq q} D_{pq}(\theta_{pq})$.
In Fig.~\ref{fig:1-UCJ-prime}, the structure of the corresponding circuit is shown.

The gate $V_\alpha$ acts on a specific pair of qubits, and corresponds to $\mathcal{R}_y(-\pi/2)$ when the pair is identified as a single IQP qubit as in Fig.~\ref{fig:1-UCJ_IQP}.
The index $\alpha$ specifies the circuit layer of 1-UCJ$_{\rm JW}^\prime$ as well as the qubit index in the corresponding IQP circuit.

$R_{n-\alpha-1, n+\alpha}(\theta)$ acts on $2\alpha+2$ qubits, among which $2\alpha$ qubits correspond to the intervening state $\ket{b}$.
Its Hamming weight is $w(b)=\alpha$, regardless of the details of the circuit, as briefly explained below.
Consider a subsystem consisting of an even number of qubits symmetrically laid on occupied and unoccupied orbitals, with a definite particle number. $V_\alpha$, and $V_\alpha^\dagger$ as well, cause the mixing between $\ket{01}$ and $\ket{10}$, but preserve the particle number in the subsystem due to their symmetric action on occupied and unoccupied orbitals (see Fig.~\ref{fig:1-UCJ-prime}).
Each $D_{pq}$ is diagonal and hence also conserves the particle number. Thus, the subsystem particle number is conserved at each step of the 1-UCJ$_{\rm JW}^\prime$ circuit. Therefore, returning to the intervening state $\ket{b}$, which is initialized with Hamming weight $\alpha$ (see Eq.~\eqref{eq:ref-state}), $w(b)=\alpha$ is maintained throughout the circuit due to the conservation of the subsystem particle number.
This ensures that the rescaling of the rotation angle in Eq.~\eqref{eq:gate-V} cancels the parity factor, such that $V_\alpha$ implements $\mathcal{R}_y(-\pi/2)$.

\subsection{
Proving 1-UCJ includes IQP
}
\label{ucj-bqp-subsection}

We finally show that for an arbitrary $n$-qubit IQP circuit, there exists an equivalent 1-UCJ$_{\rm JW}^\prime$ circuit with $2n$ qubits.

Recall that $n$-qubit IQP circuits take the form as 
\begin{align*}
\mathcal{U}_{\rm IQP}
=\mathcal{H}^{\otimes n} e^{i \mathcal{D}} \mathcal{H}^{\otimes n}.
\end{align*}
As the Pauli operator $\mathcal{Z}$ commutes with the diagonal circuit $e^{i\mathcal{D}}$, we may rewrite it as
\begin{align*}
\mathcal{U}_{\rm IQP}
&=(\mathcal{H}\mathcal{Z})^{\otimes n} e^{i \mathcal{D}} (\mathcal{Z}\mathcal{H})^{\otimes n}\notag\\
&= [\mathcal{R}_y(-\pi/2)^\dagger]^{\otimes n}e^{i \mathcal{D}}[\mathcal{R}_y(-\pi/2)]^{\otimes n},
\end{align*}
where $\mathcal{R}_y(-\pi/2) = \mathcal{ZH}$ is used.

We have already seen around Eq.~\eqref{eq:gate-V} that $\mathcal{R}_y(-\pi/2)$ can be implemented by $V_\alpha$, where 
$\alpha$ corresponds to index of the IQP qubit defined as (see also Fig.~\ref{fig:1-UCJ_IQP})
\begin{align}\label{eq:qubit-identification}
\ket{\bar{0}}_\alpha &= \ket{0}_{n-\alpha -1}\ket{1}_{n+\alpha}, \notag\\
\ket{\bar{1}}_\alpha &= \ket{1}_{n-\alpha-1}\ket{0}_{n+\alpha},
\end{align}
for $\alpha=0,1,\cdots, n-1$.

\begin{figure*}[tb]
\begin{quantikz}
n-\alpha-1~ & \gate{D_{n-\alpha-1,n-\alpha-1}(\frac{\theta}{2})} & \\
n+\alpha~\quad & \gate{D_{n+\alpha,n+\alpha}(-\frac{\theta}{2})} & 
\end{quantikz}
corresponds to 
\begin{quantikz} \\
\alpha~~ & \gate{\mathcal{R}_z(\theta)}  & \\
\end{quantikz} 
\captionof{figure}{
Construction of $\mathcal{R}_z(\theta)$ gate acting on an IQP qubit using a pair of $D_{pp}$-type operators acting on 1-UCJ$_{\rm JW}^\prime$ qubits.}
\label{fig:Rz_IQP}
\end{figure*}
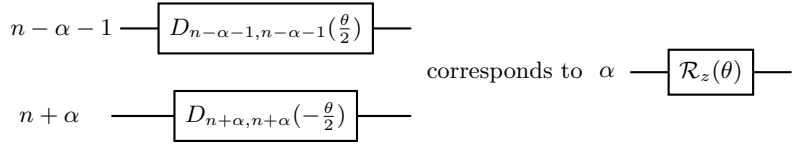
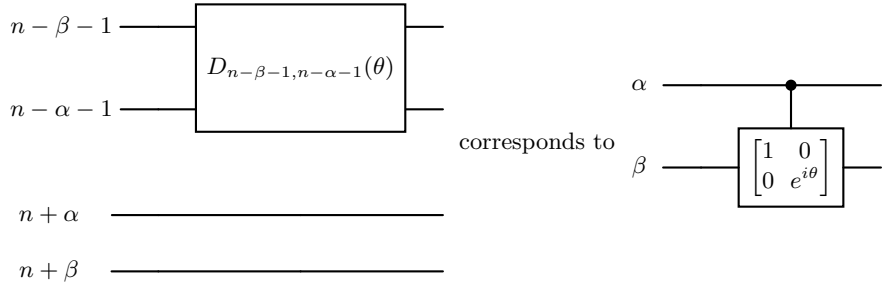
\begin{figure*}[tb]
\begin{quantikz}
n-\beta-1~ &  & \gate[wires=2]{D_{n-\beta-1,n-\alpha-1}(\theta)} & \\
n-\alpha-1~ &  &  & \\
\\
n+\alpha~\quad &  &  & \\
n+\beta~\quad &  &  &
\end{quantikz} 
corresponds to
\begin{quantikz}[transparent] \\
\alpha~~ && \ctrl{1} & \\
\beta~~ && \gate{\begin{bmatrix}
    1 & 0 \\
    0 & e^{i\theta}
\end{bmatrix}}& \\
\end{quantikz} 
\captionof{figure}{
Construction of the controlled phase shift gate acting on a pair of IQP qubits $\alpha$ and $\beta$ using a single $D_{pq}$-type operator (and identity operators)  acting on four 1-UCJ$_{\rm JW}^\prime$ qubits.
Note that in the controlled phase shift gate, the control and target qubits can be exchanged without altering the overall effect.}
\label{fig:c-phase-shift_IQP}
\end{figure*}

For the diagonal part $e^{i \mathcal{D}}$ in IQP, we assume the quadratic form
\begin{align*}
\mathcal{D} = \sum_{\alpha <\beta}w_{\alpha\beta}\mathcal{Z}_\alpha \mathcal{Z}_\beta +\sum_{\alpha}v_\alpha \mathcal{Z}_\alpha,
\end{align*}
where$w_{\alpha\beta}$ and $v_\alpha$ are real parameters. Note that restricting the summation to $\alpha < \beta$ does not affect generality, as $Z_\alpha$ and $Z_\beta$ commute.
$e^{i\mathcal{D}}$ can be rewritten as
\begin{align}\label{eq:iqp_diagonal_decomposed}
e^{i\mathcal{D}} 
&=\exp\left(-i\sum_{\alpha < \beta} w_{\alpha\beta}\right)
\exp\left( i\sum_\alpha v_\alpha^\prime \mathcal{Z}_\alpha \right) \notag\\
&\quad \times\exp\left[i\sum_{\alpha<\beta} w_{\alpha\beta}(\mathcal{I}_\alpha -\mathcal{Z}_\alpha)(\mathcal{I}_\beta -\mathcal{Z}_\beta) \right],
\end{align}
where $v_\alpha^\prime = v_\alpha +\sum_{\beta (\neq \alpha)} w_{\alpha\beta}$, with $w_{\alpha\beta}=w_{\beta\alpha}$ assumed. Ignoring the overall phase factor, this can be implemented if the rotation gate $\exp(i\theta \mathcal{Z})$, and the controlled phase shift gate $\exp[i\theta(\mathcal{I}-\mathcal{Z})\otimes(\mathcal{I}-\mathcal{Z})]$ are available.
In the following, we show that these gates can be constructed by projecting $D_{pq}$ onto IQP qubits.
The details are left to Appendix~\ref{appendix:diagonal}.

Recall the JW representation of $D_{pq}$ is $D_{pq}(\theta)=\exp [ i\theta (\frac{I_p -Z_p}{2})\otimes (\frac{I_q -Z_q}{2}) ]$ for $p\neq q$, while $D_{pp}(\theta) = \exp[i\theta (I_p - Z_p)/2]$.
It is easily observed that the action of $D_{pp}(\theta/2)D_{qq}(-\theta/2)$ on $(\ket{0_p1_q}, \ket{1_p 0_q})$, identified as a single IQP qubit $(\ket{\bar{0}},\ket{\bar{1}})$, is equivalent to $\mathcal{R}_z(\theta)=\exp(-i\theta \mathcal{Z}/2)$.

The construction of the controlled phase shift gate invokes two pairs of 1-UCJ qubits, $(\ket{0_p1_{p^\prime}},\ket{1_p 0_{p^\prime}})$ and $(\ket{0_q1_{q^\prime}},\ket{1_q 0_{q^\prime}})$.
If the first qubits $p,q$ are entangled by $D_{pq}(\theta)=\exp [ i\theta (\frac{I_p -Z_p}{2})\otimes (\frac{I_q -Z_q}{2}) ]$, then its effect on two IQP qubits is equivalent to the controlled phase shift gate $\exp[i\theta(I-\mathcal{Z})\otimes(I-\mathcal{Z})]$.
This is explicitly shown in Appendix~\ref{appendix:diagonal}.

Circuit diagrams for construction of these diagonal gates are given in Figs.~\ref{fig:Rz_IQP} and \ref{fig:c-phase-shift_IQP}.
One may think the controlled phase shift gate can act only in a limited way, that is, the control qubit index ($\alpha$ in Fig.~\ref{fig:c-phase-shift_IQP}) should be smaller than the target qubit index ($\beta$), due to the restriction $p \leq q$ imposed on $D_{pq}$.
This is not an issue since the controlled phase shift gate does not distinguish control and target qubits;
it suffices if the gate is constructed for $\alpha < \beta$, as already shown.

We have shown that the diagonal circuit $e^{i\mathcal{D}}$ in IQP can be reproduced by the operator $\prod_{p\leq q} D_{pq}(\theta_{pq})$ in 1-UCJ$_{\rm JW}^\prime$ with appropriate choice of parameters.
The HF reference state, $\ket{\psi_{\rm ref}}$ in Eq.~\eqref{eq:ref-state}, can be cast into the form of the input state for IQP, i.e., $\ket{\bar{0}^n}$ (see Fig.~\ref{fig:IQP_circuit}).
In total, it is shown that any $n$-qubit IQP circuit with a quadratic form of $\mathcal{D}$ can be emulated by a 1-UCJ$_{\rm JW}^\prime$ circuit, implying
\begin{align}
1\text{-}UCJ_{JW}^\prime \supseteq IQP.
\end{align}
As 1-UCJ$_{\rm JW}^\prime$ is a restricted class of 1-UCJ$_{\rm JW}$, then it follows
\begin{align}
1\text{-}UCJ_{JW} \supseteq IQP.
\end{align}
This is Theorem~\ref{th:our-theorem}, which is our main contribution.

Given there are IQP circuits that are classically hard to sample~\cite{iqp2010}, this in turn implies the existence of 1-UCJ circuits with the same hardness, i.e., the worst-case hardness in weak simulation.
Specifically, we have shown that the polynomial hierarchy would collapse to the third level if 1-UCJ circuits could be efficiently sampled by classical computation with multiplicative error $1 \leq c < \sqrt{2}$.
Therefore, given the collapse of the polynomial hierarchy is unlikely, it is implausible that 1-UCJ circuits are classically tractable to sample.
As a side result, we note that the following can also be shown by repeating essentially the same proof as in Ref.~\cite{iqp2010}: post-1-UCJ$_{\rm JW}=$ post-IQP $=$ post-BQP.

\section{Summary and discussion}
\label{sec:summary}

In this work, we have shown that the unitary cluster Jastrow with only a single layer (1-UCJ), a class of quantum circuits widely used in both quantum chemistry and condensed matter physics, is classically hard to sample from in the worst case. The core of the proof is to explicitly construct a circuit that can be reduced to an IQP circuit by fixing the parameters. This indicates that the polynomial hierarchy would collapse to the third level if there existed a classically efficient sampling algorithm for 1-UCJ circuits.
Since the 1-UCJ ansatz is a compact and cost-efficient variant of the unitary coupled cluster ansatz, our work marks an initial step toward establishing quantum advantage using a wide range of practically motivated circuits.
Our results have significant implications for potential quantum advantage of the quantum-selected configuration interaction method~\cite{qsci}, which employs quantum computers only for sampling.

There are numerous future directions to explore.
First, it would be crucial to investigate whether the hardness results hold for average-case instances. Considering that average-case hardness is also an important open problem for IQP circuits with 2-local interactions, this may significantly advance our understanding of the boundary between  quantum and classical instances.
Second, it is practically important to investigate whether there is advantage in noisy implementation. Envisioning that the noise rates of quantum devices become increasingly lower in the era of early fault-tolerant quantum computing, it would be crucial to investigate whether the outcome is classically tractable in the low-noise regime. Establishing such a difference would greatly strengthen the case for practical quantum advantage.
Third, classical simulatability of expectation value estimation task is an important open problem. Given that the expectation values for IQP circuits are classically simulatable~\cite{nest2011simulating}, we expect that similar results hold for 1-UCJ as well, while it remains nontrivial whether quantum advantage is gained in the UCJ ansatz with multiple layers.

\section*{Note Added}

While finalizing our manuscript for submission, we became aware of a concurrent study~\cite{leimkuhler2025exponential} posted on arXiv, which also addresses the classical hardness for sampling of quantum chemistry circuits. Among their several findings, the one most relevant to our work is that they prove the worst-case hardness of both weak and strong simulation for the unitary coupled cluster with singles and doubles (UCCSD), under the same assumption as ours, namely, that the polynomial hierarchy does not collapse. Their proof setup uses matchgate circuits with fermionic magic state inputs, which differs from ours. We emphasize that the unitary cluster Jastrow ansatz with a single layer, adopted in our work, is more cost-efficient than UCCSD, and that our result on classical hardness would have more favorable implications for near-term quantum hardware.

\section*{Acknowledgements}

A preliminary version of this work was presented as a poster at QIP 2025~\cite{HafidQIP2025}.
We thank Soumik Ghosh, Keita Kanno, and Kosuke Mitarai for fruitful discussions.
K.T. is supported by the Program for Leading Graduate Schools (MERIT-WINGS) and JST BOOST Grant Number JPMJBS2418.
N.Y. is supported by JST Grant Number JPMJPF2221, JST CREST Grant Number JPMJCR23I4, IBM Quantum, JST ASPIRE Grant Number JPMJAP2316, JST ERATO Grant Number JPMJER2302, and Institute of AI and Beyond of the University of Tokyo.

\appendix

\section{
Constructing diagonal gates of IQP from 1-UCJ
}\label{appendix:diagonal}

In this Appendix, we show that the diagonal part $e^{i\mathcal{D}}$ in an IQP circuit $\mathcal{H}^{\otimes n}e^{i\mathcal{D}}\mathcal{H}^{\otimes n}$ can be constructed from $D_{pq}(\theta)$, a diagonal component in a 1-UCJ (or more precisely, 1-UCJ$_{\rm JW}^\prime$) circuit defined in Eq.~\eqref{eq:1-ucj-prime-def}.
Note we use calligraphic characters such as $\mathcal{H}$ to denote operators in IQP, while usual ones such as $H$ for 1-UCJ$_{\rm JW}^\prime$.

In this paper, we assume the following quadratic form for $\mathcal{D}$:
\begin{align}
\mathcal{D} = \sum_{\alpha <\beta}w_{\alpha\beta}\mathcal{Z}_\alpha \mathcal{Z}_\beta +\sum_{\alpha}v_\alpha \mathcal{Z}_\alpha,
\end{align}
where $w_{\alpha\beta}$ and $v_\alpha$ are real parameters. This can be rewritten as
\begin{align}
e^{i\mathcal{D}} 
&=\exp\left(-i\sum_{\alpha < \beta} w_{\alpha\beta}\right)
\exp\left( i\sum_\alpha v_\alpha^\prime \mathcal{Z}_\alpha \right) \notag\\
&\quad \times\exp\left[i\sum_{\alpha <\beta} w_{\alpha\beta}(\mathcal{I}_\alpha -\mathcal{Z}_\alpha)(\mathcal{I}_\beta -\mathcal{Z}_\beta) \right],
\end{align}
where $v_\alpha^\prime=v_\alpha+\sum_{\beta (\neq \alpha)}w_{\alpha\beta}$, with $w_{\alpha\beta}=w_{\beta\alpha}$ assumed.
Here, \(\exp\left(-i\sum_{\alpha<\beta} w_{\alpha\beta}\right)\) is a global phase factor, which can be ignored.
Meanwhile, \(\exp\left[i w_{\alpha\beta}(\mathcal{I}_\alpha-\mathcal{Z}_\alpha)(\mathcal{I}_\beta-\mathcal{Z}_\beta) \right]\) corresponds to a controlled phase shift gate, which we denote as \(\mathcal{CP}_{\alpha\beta}(\theta)\) with control and target indices $\alpha,\beta$, and \(\exp\left( i v_\alpha^\prime \mathcal{Z}_\alpha \right)\) corresponds to a rotation gate about the $z$-axis, \(\mathcal{R}_z(\theta)\).
The gates are defined as $\mathcal{R}_z(\theta)=e^{-i\theta \mathcal{Z}/2}$ and $\mathcal{CP}_{\alpha\beta}(\theta)=\exp\left[ i\theta\left(\frac{\mathcal{I}_\alpha - \mathcal{Z}_\alpha}{2}\right)\otimes \left(\frac{\mathcal{I}_\beta - \mathcal{Z}_\beta}{2}\right) \right]$ for $\alpha\neq \beta$. The matrix representation of the latter is given as
\begin{align}\label{eq:CP_IQP_matrix}
\mathcal{CP}_{\alpha\beta}(\theta) = \begin{pmatrix}
 1 & 0 & 0 & 0 \\
 0 & 1 & 0 & 0 \\
 0 & 0 & 1 & 0 \\
 0 & 0 & 0 & e^{i\theta} \\
\end{pmatrix},
\end{align}
for $(\ket{00}, \ket{01}, \ket{10}, \ket{11})$.
Note that $\mathcal{CP}_{\alpha\beta}$ is symmetric with respect to $\alpha$ and $\beta$, and does not distinguish the control and target qubits.

In the following, we show that $\mathcal{R}_z(\theta)$ and $\mathcal{CP}_{\alpha\beta}(\theta)$ can be constructed from products of $D_{pq}(\theta)$,
\begin{align}
\prod_{p\leq q} D_{pq}(\theta_{pq}),
\end{align}
contained in a 1-UCJ$_{\rm JW}^\prime$ circuit of Eq.~\eqref{eq:1-ucj-prime-def}.
With the Jordan-Wigner transformation, it is given as
\begin{align*}
&D_{pq}(\theta) \\ 
&= \begin{dcases}
    \exp\left[ i\theta \left(\frac{I_p -Z_p}{2}\right)\otimes\left(\frac{I_q -Z_q}{2}\right) \right] & \text{for $p\neq q$,} \\
    \exp[i\theta (I_p - Z_p)/2] & \text{for $p=q$.}
\end{dcases}
\end{align*}
A single qubit in IQP $(\ket{\bar{0}},\ket{\bar{1}})$ is constructed from two qubits, or a particle number conserving subspace $(\ket{01}, \ket{10})$, in 1-UCJ$_{\rm JW}^\prime$ following the identification (see also Fig.~\ref{fig:1-UCJ_IQP}):
\begin{align}\label{eq:qubit-identification_app}
\ket{\bar{0}}_\alpha &= \ket{0}_{n-\alpha -1}\ket{1}_{n+\alpha}, \notag\\
\ket{\bar{1}}_\alpha &= \ket{1}_{n-\alpha-1}\ket{0}_{n+\alpha},
\end{align}
for $\alpha=0,1,\cdots, n-1$.
We construct $\mathcal{R}_z$ and $\mathcal{CP}$ on such IQP qubits, in turn.

The construction of $\mathcal{R}_z$ is simple.
Define the operator $F_\alpha(\theta)$ acting on a pair of $n-\alpha-1$-th and $n+\alpha$-th as
\begin{align}
F_{\alpha}(\theta) = D_{n-\alpha-1,n-\alpha-1} \left(\frac{\theta}{2} \right) D_{n+\alpha,n+\alpha}\left( -\frac{\theta}{2} \right).
\end{align}
Then, we see that its action is closed within the particle number conserving subspace as
\begin{align}
F_{\alpha}(\theta)\ket{0}_{n-\alpha-1}\ket{1}_{n+\alpha} &= e^{-i\frac{\theta}{2}}\ket{0}_{n-\alpha-1}\ket{1}_{n+\alpha},\nonumber\\
F_{\alpha}(\theta)\ket{1}_{n-\alpha-0}\ket{0}_{n+\alpha} &= e^{i\frac{\theta}{2}}\ket{1}_{n-\alpha-1}\ket{0}_{n+\alpha}.
\end{align}
Restricting $F_\alpha(\theta)$ within this subspace and renaming it as $\mathcal{R}_z(\theta)$, its action on the IQP qubit defined in Eq.~\eqref{eq:qubit-identification_app} is given as
\begin{align}
\mathcal{R}_{z}(\theta)\ket{\bar{0}}_{\alpha} 
&= e^{-i\frac{\theta}{2}}\ket{\bar{0}}_{\alpha}, \notag\\
\mathcal{R}_{z}(\theta)\ket{\bar{1}}_{\alpha} 
&= e^{i\frac{\theta}{2}}\ket{\bar{1}}_{\alpha}.
\end{align}
This actually corresponds to the rotation about the $z$-axis for the $\alpha$-th qubit in the The corresponding circuit diagram is given in Fig.~\ref{fig:Rz_IQP}.

The construction of the controlled phase shift gate $\mathcal{CP}_{\alpha\beta}$ invokes four 1-UCJ qubits, which are supposed to be paired in the pairwise particle number conserving manner as $(\ket{0}_{n-\alpha-1}\ket{1}_{n+\alpha}, \ket{1}_{n-\alpha-1}\ket{0}_{n+\alpha})$ and $(\ket{0}_{n-\beta-1}\ket{1}_{n+\beta}, \ket{1}_{n-\beta-1}\ket{0}_{n+\beta})$ for $\alpha\neq \beta$.
Define the operator $D^\prime_{\alpha\beta}(\theta)$ acting on the four qubits as
\begin{align}
D^\prime_{\alpha\beta}(\theta) = D_{n-\beta-1,n-\alpha-1}(\theta) \otimes I_{n+\alpha} \otimes I_{n+\beta},
\end{align}
for $0 \le \alpha < \beta \leq n-1$.
Then, we observe that its action is closed within the four-dimensional pairwise particle number conserving subspace as
\begin{align}
&D^\prime_{\alpha\beta}(\theta)\ket{0}_{n-\beta-1}\ket{0}_{n-\alpha-1}\ket{1}_{n+\alpha}\ket{1}_{n+\beta} \nonumber\\
&= \ket{0}_{n-\beta-1}\ket{0}_{n-\alpha-1}\ket{1}_{n+\alpha}\ket{1}_{n+\beta},\nonumber\\
&D^\prime_{\alpha\beta}(\theta)\ket{1}_{n-\beta-1}\ket{0}_{n-\alpha-1}\ket{1}_{n+\alpha}\ket{0}_{n+\beta} \nonumber\\
&= \ket{1}_{n-\beta-1}\ket{0}_{n-\alpha-1}\ket{1}_{n+\alpha}\ket{0}_{n+\beta},\nonumber\\
&D^\prime_{\alpha\beta}(\theta)\ket{0}_{n-\beta-1}\ket{1}_{n-\alpha-1}\ket{0}_{n+\alpha}\ket{1}_{n+\beta} \nonumber\\
&= \ket{0}_{n-\beta-1}\ket{1}_{n-\alpha-1}\ket{0}_{n+\alpha}\ket{1}_{n+\beta},\nonumber\\
&D^\prime_{\alpha\beta}(\theta)\ket{1}_{n-\beta-1}\ket{1}_{n-\alpha-1}\ket{0}_{n+\alpha}\ket{0}_{n+\beta} \nonumber\\
&= e^{i\theta}\ket{1}_{n-\beta-1}\ket{1}_{n-\alpha-1}\ket{0}_{n+\alpha}\ket{0}_{n+\beta} .
\end{align}
Restricting $D^\prime_{\alpha\beta}(\theta)$ within this subspace and renaming it as $\mathcal{CP}_{\alpha\beta}(\theta)$, its action on the pair of IQP qubits defined in Eq.~\eqref{eq:qubit-identification_app} is given as
\begin{align}
\mathcal{CP}_{\alpha\beta}(\theta)\ket{\bar{0}}_{\alpha}\ket{\bar{0}}_{\beta} &= \ket{\bar{0}}_{\alpha}\ket{\bar{0}}_{\beta},\nonumber\\[1mm]
\mathcal{CP}_{\alpha\beta}(\theta)\ket{\bar{0}}_{\alpha}\ket{\bar{1}}_{\beta} &= \ket{\bar{0}}_{\alpha}\ket{\bar{1}}_{\beta},\nonumber\\[1mm]
\mathcal{CP}_{\alpha\beta}(\theta)\ket{\bar{1}}_{\alpha}\ket{\bar{0}}_{\beta} &= \ket{\bar{1}}_{\alpha}\ket{\bar{0}}_{\beta},\nonumber\\[1mm]
\mathcal{CP}_{\alpha\beta}(\theta)\ket{\bar{1}}_{\alpha}\ket{\bar{1}}_{\beta} &= e^{i\theta}\ket{\bar{1}}_{\alpha}\ket{\bar{1}}_{\beta}.
\end{align}
This is the right operation of the controlled phase shift gate on qubits $\alpha$ and $\beta$, as given in Eq.~\eqref{eq:CP_IQP_matrix}.
The corresponding circuit diagram is given in Fig.~\ref{fig:c-phase-shift_IQP}.

Therefore, we have shown that the diagonal part $e^{i\mathcal{D}}$ of any IQP circuit with $\mathcal{D}$ quadratic in the Pauli $\mathcal{Z}$ can be emulated by a 1-UCJ$_{\rm JW}^\prime$ utilizing appropriate sequences of its diagonal elements $D_{pq}$.

\bibliography{references}

\end{document}